\documentclass[onecolumn,floatfix,superscriptaddress,showpacs]{revtex4}
\usepackage{graphicx}
\usepackage{helvet}
\usepackage{graphicx}
\usepackage{color}
\usepackage{psfrag}
\usepackage{epsfig}
\usepackage{bm}
\newcommand{\ket}[1]{| #1 \rangle}
\newcommand{\bra}[1]{\langle #1 |}
\newcommand{\rb}[1]{\left( #1 \right)}

\newcommand{\beq}{\begin{eqnarray}}
\newcommand{\eeq}{\end{eqnarray}}

\newcommand{\stateuu}{| \uparrow \uparrow \rangle}
\newcommand{\statedd}{| \downarrow \downarrow \rangle}
\newcommand{\stateud}{| \uparrow \downarrow \rangle}
\newcommand{\statedu}{| \downarrow \uparrow \rangle}

\begin{document}
{
\title{\Large Measuring the squared magnetisation of a molecule}
\author{Bj\"orn Michaelis}
\address{Instituut-Lorentz, Universiteit Leiden, \\
P.O. Box 9506, 2300 RA Leiden, The Netherlands}
\begin{abstract}
It is proposed to shoot a molecule through a pipe of mesoscopic scale.
The molecule should effectifely carry two spins and antiferromagnetic chains should be embedded in the wall of the tube. 
If the momentum is detected before and behind the tube, a projection of the molecular spins $\vec{\sigma}_1$ and $\vec{\sigma}_2$
takes place. If the molecule did not loose momentum inside the tube, it has a vanishing squared total spin in the direction of its anisotropy axis $\vec{l}$.
Whereas $[\vec{l}(\vec{\sigma}_1+\vec{\sigma}_2)]^2=1$ if it did so.
\end{abstract}
\maketitle
\large
\section{introduction}
It is one of the most broadcasted understandings from Quantum Mechanics, that a complete measurement - e.g. the simultaneous, arbitrary
well detection of the position and momentum of a particle - is impossible.
Thus one could be lead to the opinion,
that quantum mechanics provides less principle possibilities than a classical model. That this is not true can be seen the most 
clean in the ideas of quantum computation \cite{qinf1,qinf2}. To push these formal ideas into realization, one has to get control over the
way the wavefunction gets processed.\\
It has been shown that an important elementary step in this processing is an quantum mechanically incomplete measurement - but not in the sense 
above. It is easier to understand the distinction by means of discrete particle properties than with momentum and position. We use the
working horse hilbert space of two spin half particles for it. The spin part of the wavefunction can be in a superposition of two spins up,
$\stateuu$, two spins down, $\statedd$, and the mixed versions $\stateud,\statedu$
\footnote{all we say in this paper can be generalised in the following way: $\stateud$ and $\statedu$ can be substituted by all linear combinations of them,
thus e.g the singlet and the triplet with vanishing magnetisation}.
 A quantum mechanically complete measurement
would now provide  to the measurer one out of four possible results. The pointer of the detector shows e.g. $1$ for $\stateuu$, $2$ 
for $\statedd$ etc.. After that, the wavefunction is left in the detected state. It was complete, because the measurer could not
have obtained more information about the particle.
Determining the total magnetisation  in z direction, thus $1,0$ or $-1$, would e.g. be an quantum mechanically incomplete 
measurement. The difference is, that for the outcome $0$, the wavefunction is still in an unknown superposition of $\stateud$ and 
$\statedu$.\\
The measurement that was proven to be useful for quantum computation is another one \cite{parityidea,parityidea2}. It is the ${\cal S}_{tot}^2\equiv[\vec{l}(\vec{\sigma}_1+\vec{\sigma}_2)]^2$
(or more generally the parity-) measurement along some axis $\vec{l}$. In each of the two possible outcomes, the wavefunction is still 
left in a superposition.
There are several proposals for this kind of measurement\cite{parity1,parity2,parity3,parity4}. All of them have in common that the spins (which contain the quantum 
information) are localised in quantum dots. A macroscopic current can pass by, acts as detector\cite{field} and projects the dots through coulomb 
interaction into
the ${\cal S}_{tot}^2=0$ or ${\cal S}_{tot}^2=1$ space. Our paper complements those proposals in the sense that the information carrier is moving and the
detector is stationary.

\section{setup and requirements  } 
\fontseries{16}
The here proposed experimental setup is sketched in figure \ref{fig1}. It consists of four distinctive  parts. 
The magnetic oven provides a beam of molecules, that are each in an unknown superposition of their internal magnetic
states $\stateud, \statedu, \statedd, \stateuu$. We write them in the laboratory fixed basis in z-direction.
 They are assumed to leave the oven non entangled  
with any dynamical degree left in the oven. They are further in a product state with the centre of mass
of the molecule.
The splitting of their energies $\Omega_{\alpha}$ is besides the kinetic energy of the molecule itself large compared
 to all energy transfers in the further
discussion.
Once a while some molecule may pass the velocity selecting wheels and a state $\phi_{R_A,k_A}(k)$ with mean momentum $\hbar k_A$ is 
prepared at time $t_A$ at position $R_A$. It passes through the pipe, can excite the antiferromagnetic
chains embedded in its wall and becomes ideally detected
at time $t_B$ in a wavepacket $\phi_{R_Bk_B}(k)$ characterised by $k_B,R_B$ and  $\sigma_B$.
\beq
\phi_{X}(k)&=&(2\pi\sigma_x)^{(1/4)}exp[-\sigma_x^2(k-k_x)^2+i(k_x-k)R_x]\nonumber \\
&&\qquad\qquad\qquad\qquad\qquad\qquad\qquad   X=k_x,R_x
\eeq
The nature of the preparation and the detection determines $\sigma_{A,B}$. If one uses e.g laser barriers to determine the time of flight
, different light wavelengths can project the molecule into wavefunctions of different shape.\\  
We will see in the following, that the spin channels $\stateud,\statedu$ can 
not undergo inelastic scattering, whereas the channels $\stateuu,\statedd$ can loose some
of their kinetic energy and excite the antiferromagnet. An {\it almost} good setup
 for a parity measurement requires for the transmission amplitudes 
\beq
|T_{el}^{x}|&\ll &|T_{in}^{x}|\qquad\mbox{for } x=\stateuu,\statedd .
\label{almostgood}
\eeq
The observed momenta $\hbar k_B$ for the inelastic scattered molecules are expected to fulfil 
\beq
(k_A-k_B)^2&\gg &\frac{1}{\sigma_A^2}+\frac{1}{\sigma_B^2}.
\eeq 
Thus a momentum detection projects into the ${\cal S}_{tot}^2=1$ space , if the momentum change is much more than $\hbar\sqrt{1/\sigma_A^2+1/\sigma_B^2}$ and to ${\cal S}_{tot}^2=0$, if it
isn`t.
The word {\it almost} is referring to the projection to ${\cal S}_{tot}^2=1$. Because the reacting channels can each excite different states in the antiferromagnet,
 there could in principle
be situations in which the further processing of the molecule after the detection at $R_B$ can not be explained by just looking at a superposition of 
the $\stateuu,\statedd$ states. In the worst case, the spin degrees are entangled like $\stateuu\ket{EX_\uparrow}+\statedd\ket{EX_\downarrow}$ 
and the two excitations inside the tube, $\ket{EX_{\uparrow/\downarrow}}$, decay each in independent processes to orthogonal  bath states (e.g.` lost phonons`). The remaining
molecule spin states behave then just like an incoherent mixture \ref{encoding}. But a quantitative discussion of that point is deferred to chapter 
\ref{chptnoencoding} and we first give the inequation (\ref{almostgood}) a microscopic foundation 
and elaborate how to design the experiment to fulfil it.\\
\begin{figure}[t]
\begin{center}
\epsfig{file=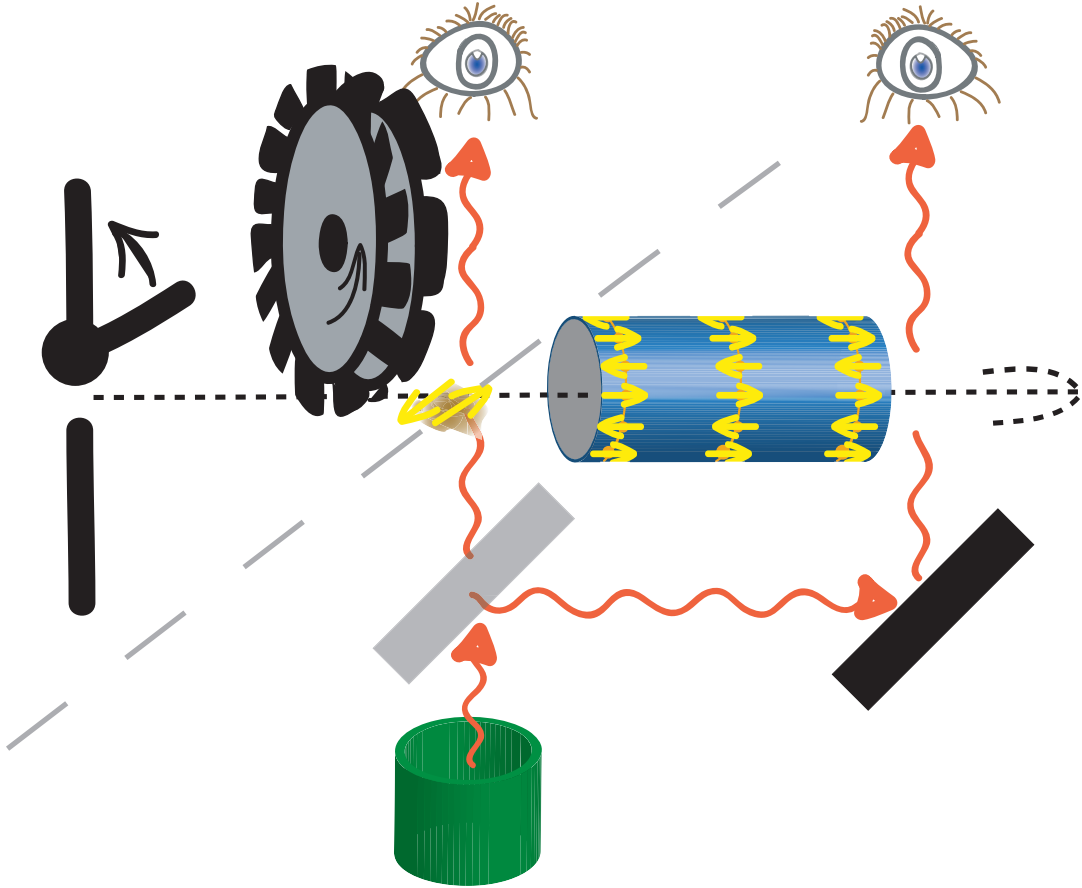, angle=0,clip=true,width=0.8\linewidth}
\end{center}
  \caption{Sketch of the proposed squared magnetisation measurement. A molecule with spin quantisation axis $\vec{l}$ (along the large spaced dashed line) is emitted
 from the source at the left. Its spins $\vec{\sigma}_1,\vec{\sigma}_2$ are in an arbitrary superposition. It has to 
pass velocity selecting wheels and propagates in z direction (along the dashed arrow). Then a first laser beam detects it at time $t_A$ 
at position $R_A$. It passes
a tube with antiferromagnetic chains in its wall. Some part of the molecular wavefunction leaves a trace in the tube by exciting
spin waves and looses kinetic energy. Then there is at time $t_B$ a space-momentum detection at $R_B$.
All molecules which lost sufficient momentum got projected into the subspace with  $[\vec{l}(\vec{\sigma}_1+\vec{\sigma}_2)]^2=1$, the others have 
$ [\vec{l}(\vec{\sigma}_1+\vec{\sigma}_2)]^2=0$.}
  \label{fig1}
\end{figure}

\section{expectation values and transmission amplitudes}
The expectation values to detect in the channel $x$ the molecule with momentum $\hbar k_B$ under the condition,
that the target stays in the groundstate or the n-th chain is left in the $\lambda$-th excited state, are related to the transmission amplitudes
by
\beq
&&|T_{el}^{x}|=\left| \int~ dkdk` \phi_B(k)\phi_A^*(k`)g^0_{k,k`}(t_B-t_A,x)\right| \\
&&g^{0}_{k,k`}(t,x)=\bra{0}a_kU^{(x)}(t)a^{\dagger}_{k'}  \ket{0}\\
&&|T_{in}^x|=\sum_{\lambda,n}\left| \int dkdk' \phi_B(k)\phi_A^*(k')g^{\lambda,n}_{k,k'}(t_B-t_A,x)\right| \\
\label{inelastictransmission}
&&g^{\lambda,n}_{k,k'}(t,x)=\bra{0}a_k\gamma_{\lambda,n}U^{(x)}(t)a^{\dagger}_{k'}  \ket{0}\\
&&\qquad \qquad\qquad \qquad\qquad x=\stateuu,\statedd,\stateud,\statedu\nonumber
\eeq
$\gamma^{\dagger}_{\lambda,n}$ create magnon excitations in the antiferromagnet, they are defined in
equation (\ref{fourmodes}).
 $a_k^{\dagger}$ is a creation
operator, it is related to the field operator of the centre of mass coordinate of the molecule
\beq
&&\ket{R}\bra{R}=\int dkdk' e^{i(k-k')R}a^{\dagger}_ka_{k'}.
\label{cmoperators}
\eeq
Without loss of generality can it be chosen to be bosonic because molecule molecule interference is of no relevance here. 
The time evolution operator
\beq
U^{(x)}(t)&=&exp(-it(H_0^{(x)}+V_{dip}^{(x)})/\hbar)
\eeq
depends parametrically on the index of the quenched molecular spin state $x$. In the hamiltonian 
\beq
H_0^{(x)}&=&
\Omega_{x}
+\int dk~ E(k)a_k^{\dagger}a_k
+W\sum_{\lambda,n}\gamma_{\lambda,n}^{\dagger}\gamma_{\lambda,n}\nonumber \\
&&\qquad\qquad\qquad\qquad  E(k)=\hbar^2k^2/2m  
\eeq
are only the chain degrees included to which the molecule spins couple by the magnetic dipole dipole interaction
\beq
V_{dip}^{(x)}&=\Gamma_x\sum_{\lambda,n}&\int dk~dk`~a_k^{\dagger}a_{k`}\gamma_{\lambda,n}^{\dagger}V_{tot}^{\lambda,n}(k-k`) \nonumber \\
&&\qquad \Gamma_{\{\uparrow\uparrow,\downarrow\downarrow,\uparrow\downarrow,\downarrow\uparrow\}}=\{1,-1,0,0\} .
\eeq
Appendix B derives this expression and shows that $W$ is the magnon energy gap at long wavelengths. It will be of some help to split
the potential into a part that is symmetric and antisymmetric under space inversion $V_{tot}(q)=V_{s}(q)+V_{a}(q)$. 
Only one magnon excitations are considered,
equivalent with the Born Approximation. One can make use of the complex variables $\mu_k,\xi_{\lambda,n}$, 
which label coherent states on top of the molecule-magnon vacuum $\ket{0}$,
\beq
\ket{\mu}=exp\left(\sum_k \mu_ka_k^{\dagger}\right)\ket{0} , & &\ket{\xi}=exp\left(\sum_{\lambda,n}\xi_{\lambda,n}\gamma_{n,i}^{\dagger}\right)\ket{0}.
\eeq
To perform the expansion to first order in the reaction probability of a single chain 
\beq
 g^0_{k,k'}(t,x)&=&\langle \mu_k(t)\mu_{k`}^*(0)\rangle_0=\delta_{k,k`}G(k,t)F(t)\\
 g^{\lambda,n}_{k,k'}(t,x)&=&\frac{-i\Gamma_x}{\hbar}\int dq~dq`~d\tau  V_{tot}^{\lambda,n}(q-q`) \nonumber \\
&&\times \langle\mu_k(t)\mu_{q`}^*(\tau)\mu_q^*(\tau)\mu_q^*(0)\xi_{\lambda,n}(t)\xi_{\lambda,n}^*(\tau)\rangle_0 \nonumber \\
 \eeq
one uses the Wick theorem to express the path integrals \cite{NegeleOrland}
\beq 
 \langle .. \rangle_0&=&\int D[\mu\mu^*\xi\xi^*]...
exp\left[
\int dt\left(\sum_{\lambda,n}F^{-1}(t)\xi_{\lambda,n}(t)\xi_{\lambda,n}^*(t)\right.\right. \nonumber \\
&&\left.\left.+\int dk G^{-1}(k,t)\mu_k(t)\mu_k^*(t)\right)
\right] 
\eeq
in terms of $G(t,k)=exp(-i\hbar k^2t/2m)\Theta(t)$, $F(t)=exp(-iWt/\hbar)\Theta(t)$. $\Theta(..)$ is the unit step function. 

The elastic contribution describes just the projection onto the free propagating wavepacket
\beq
 |T^{x}_{el}|=\chi(t_B-t_A)^{-1/4 }
exp&&\left[
\tiny-(k_A-k_B)^2\left(\frac{1}{\sigma_A^2}+\frac{1}{\sigma_B^2}\right)^{-1}\right. \nonumber \\
&&\left.-\frac{  ({\cal R}_{el}- {\cal V}_{el}(t_B-t_A))^2  }{   4(\sigma_A^2+\sigma_B^2)\chi(t_B-t_A)    }~\right] \nonumber 
\eeq
\beq
 {\cal R}_{el}=R_B-R_A~~\qquad&&~{\cal V}_{el}=\frac{\hbar}{m}\frac{k_A\sigma_A^2+k_B\sigma_B^2}{\sigma_A^2+\sigma_B^2}\nonumber \\
\chi(t)=1+t^2/t_{disp}^2 \qquad&& t_{disp}=2\sqrt{2}m(\sigma_A^2+\sigma_B^2)/\hbar
\label{elastictransmission2}
\eeq 
and leads for all spin channels to a detected momentum that differs only through the detection and preparation process itself from the initialised
momentum. The detection times are in the following assumed to be always much smaller than the generalised dispersion time $t_{disp}$, so, that
the molecule is at detection still in the regime of corpuscular behaviour.  \\
The propagation kernel for the inelastic contribution depends on the lower(upper) momentum cut off $\hbar Q_{<(>)}$
\beq
 g_{k,k'}^{\lambda,n}(t,x)&=&-i\delta_{k,k`}V_s^{\lambda,n}(0)G(k,t)(Q_>-Q_<)\Gamma_x\int dt` F(t-t`)/\hbar \nonumber \\
&&-iV_{tot}^{\lambda,n}(k-k`)\Gamma_x\int d\tau G(k,t-t`)G(k`,\tau)F(t-t`)/\hbar \nonumber \\
&=&
  V_s^{\lambda,n}(0)(Q_>-Q_<)\Gamma_x W^{-1}(1-e^{-iWt/\hbar})e^{-iE(k)t/\hbar}\delta_{k,k`}\nonumber  \\
&&~~+
V_{tot}^{\lambda,n}(k-k`)\Gamma_x
\frac{e^{-iE(k`)t/\hbar}-e^{-i(E(k)+W)t/\hbar}}{E(k`)-E(k)-W}
\eeq
If one inserts this into equation (\ref{inelastictransmission}), the first part gives rise to the same $k_B$ distribution as in equation 
(\ref{elastictransmission2}). Thus such a momentum detection wouldn`t reveal the excitations inside the tube. We shouldn`t worry about
the fact that it even 
diverges if one relaxes the cut offs. Including an infinite momentum spectrum is for the wavepacket and the magnetic field
anyhow unphysical.
We simplify for large enough detection times,
$\hbar \ll Min\left(E(Q_<)+W-E(Q_>),W\right)(t_B-t_A)$,
\beq
 |T_{in}^x|&=&\left|\sum_{\lambda,n}\frac{V_s^{\lambda,n}(0)(Q_>-Q_<)}{W}\left(1-e^{-iWt/\hbar}\right)|T_{el}^x|e^{-i\varphi_{in}}\right.\nonumber \\
&&+\int dkdk`\phi_B(k)V_{tot}^{\lambda,n}(k-k`)exp\left[-i(E(k)+E(k`)+W)(t_B-t_A)/(2\hbar)\right] \nonumber\\
&&~~\times 
\left.\frac{  2~sin\left[ (E(k')-E(k)-W)(t_B-t_A)/2\hbar \right]  }  {  E(k)-E(k`)-W  }
\phi_A^*(k`)e^{i\psi_{in}} \right|,~~x=\stateuu,\statedd\nonumber \\
\label{goodandbad}
\eeq
by substituting $\frac{sin((t_B-t_A)..)}{..}$ with a proper normalised step function between $-\pi/(t_B-t_A)$ and $\pi/(t_B-t_A)$.
Time dependent phases, $\varphi_{in},\psi_{in}$, appear. But they are independent of $\lambda$ and $n$ and will through the following discussion become irrelevant.
We are left with the task to perform the integral
\beq 
\quad\int dk V_{tot}^{\lambda,n}(k-\sqrt{k^2+Q^2})
\phi_B(k)
\case{exp\left[-iE(k)(t_B-t_A)/\hbar\right]}{\sqrt{k^2+Q^2}}
\phi_A^*(\sqrt{k^2+Q^2})
\label{int1}
\eeq
using the notation $Q^2=2mW/\hbar^2$. All the Besselfunctions in equations (\ref{Vasym},\ref{Vsym}) behave for large momenta like $exp(-|qd|)$. 
Thus if one wants to take full advantage
of the potential, one obtains an upper limit for the tube diameter 
\beq
\frac{1}{d}>k_B\left(1-\sqrt{1+\frac{Q^2}{k_B^2}}\right)+\sigma_B^{-1}\left(1-\frac{1}{\sqrt{1+\frac{Q^2}{k_B^2}}}\right)
\eeq
which can be ignored for
 $k_B^2 > \frac{Q^2}{1-(k_B\sigma_B)^{-2}}$.
How to calculate the integral depends essentially on the ratio of $\sigma_A$ and $\sigma_B$. 
Thus
the approximate gaussian contribution around $k=\sqrt{k_A^2-Q^2}$ and $k=k_B$ gives in the limiting cases 
\beq
\left(\frac{2\sigma_{A}\sigma_B}{\sigma_{AB}^2+\sigma_B^2}\right)^{\frac{1}{2}}
exp\left[
-{\cal Q}_{in}^2\left(  \frac{1}{\sigma_{AB}^2}  +  \frac{1}{\sigma_B^2}    \right)^{-1} 
-\frac{  \left(   {\cal R}_{in}-{\cal V}_{in}(t_B-t_A)\right)^2  }{ 4(\sigma_{AB}^2+\sigma_B^2)  }
 \right]\nonumber\\
\qquad \times 
\left\{  
\begin{array}{lr}
\frac{V_{tot}^{\lambda,n}(\sqrt{k_A^2-Q^2}-k_A)}{k_A}&{\rm if}\;\;\sigma_A\gg\sigma_B\\
\\
\frac{V_{tot}^{\lambda,n}(k_B-\sqrt{k_B^2+Q^2})}{\sqrt{k_B^2+Q^2}}&{\rm if}\;\;\sigma_B\gg\sigma_A\\
\end{array}
\right. 
\qquad(\ref{int1}')\nonumber
\eeq
with the abbreviations 
\beq
\qquad
\begin{array}{ll}
{\cal Q}_{in}=k_B-\sqrt{k_A^2-Q^2} \qquad &
{\cal V}_{in}=
\frac{\hbar}{m}\sqrt{k_A^2-Q^2}
\\
{\cal R}_{in}=
R_B-R_A\left(1-Q^2/k_A^2\right)^{1/2} 
&
\sigma_{AB}=
\sigma_A\left(1-Q^2/k_A^2\right)^{1/2} 
\end{array}
\quad(\ref{int1}'')\nonumber
\eeq
for  $\sigma_A\gg\sigma_B$
and
\beq
\begin{array}{ll}
{\cal Q}_{in}=\left(k_A-\sqrt{k_B^2+Q^2}\right)(1+Q^2/k_B^2)^{1/2}  &
\sigma_{AB}=
\sigma_A\left(1+Q^2/k_B^2\right)^{-1/2} 
\\
{\cal R}_{in}=
R_B-R_A\left(1+Q^2/k_B^2\right)^{-1/2} 
&
{\cal V}_{in}=
\frac{\hbar}{m}k_B
\qquad\qquad(\ref{int1}''')
\end{array}
\nonumber
\eeq
in the opposite limit.
If one ignores, that in equation (\ref{mainresult1}) the sums over $\lambda,n$ are inside the $|..|$ signs, the most probable momentum to detect is in both cases 
for these inelastic scattering paths 
$\hbar(k_A^2-Q^2)^{-1/2}$. But what is now the effect of taking the sum first ? Important is the sum over the z-positions
of the different spin rings $r_n$.One gets $|T_{in}^x|\propto\sum_n exp(-i(k_A-\sqrt{k_A^2-Q^2})r_n )$. If there are many
spin rings these phases would usually average out each other and we don`t find inelastic scattering at all. For this reason we have to
design the distances of the rings so that $r_n=2\pi{\cal N}(n)/(k_A-\sqrt{k_A^2-Q^2})$, where $\cal{N}$(..) gives only integer numbers.
This resembles the idea of multilayered mirrors which achieve very high reflection for a certain wavelength of light.
 Thus if one is able to design the positions of the rings in the prescribed way, 
then one can improve as well the accuracy of the momentum transfer.\\ 

Equations (\ref{goodandbad},\ref{int1}) offer two strategies two diminish the inelastic contributions that would lead to measurements of
$k_B\approx k_A$, $t_B\approx t_A+{\cal R}_{el}{\cal V}^{-1}_{el}$. 
If one is able to keep  the spin quench axis of the molecule
along its flight direction ($\beta=\pi/2$ in equations (\ref{Vasym},\ref{Vsym})), the symmetric part of the interaction $V_s$ vanishes. We insert
this condition into the requirement (\ref{almostgood}) at the optimal measurement values. We can further assume that the tube is much
shorter than its distances to the locations of preparation and detection and obtain our main result 
\beq
1
&\ll&
4\mu_0\mu_e^2\frac{M\sqrt{SN}{\cal F}(W)}{d^2v_A\hbar}\left[\frac{\sigma_A\sigma_B}{(1-W/E(k_A))\sigma_A^2+\sigma_B^2}\right]^{1/2}
\label{mainresult1}
\eeq
as an estimate for the minimal number of spin rings $M$. The function 
\beq
{\cal F}(W)=exp
\left[
\frac{1}{2}
arctanh
\left( 
\frac{1}{1+\frac{W}{2JS}}
\right)
\right]
\eeq
goes monotonously from 1 at $W/SJ\rightarrow\infty$ to $\infty$ at $W=0$. It may be tempting just to decrease
$W$, so that M doesn`t have to be very large. But the validity of the Born approximation requires that the detection
probability of a single chain is small. The opposite of condition (\ref{mainresult1}) has to be true for $M=1$,thus it gives
 a lower limit for $W$ as well.\\
In the realistic situation $J=W$, $S=1/2$, and $\sigma_B\gg\sigma_A$, these requirements can be written as
\beq
&v_{limit}\ll v_A \ll Mv_{limit}\quad \mbox{with } v_{limit}\equiv108\sqrt{\frac{\sigma_A\rho}{\sigma_B d^3}}& \nonumber \\
& v_A=439\sqrt{\frac{W}{\epsilon(2-\epsilon)m}}&
\label{inequationwithnumbers}
\eeq
, where we introduced the relative momentum loss $\epsilon$. The initial speed $v_A$ and $v_{limit}$ are given in $m/s$.  The mass is
given in  atomic
mass units (A.M.U.) and $W$ in its typical units of $meV$. $\rho$ is the distance between two spins in a chain and, like the other
lengthscales $d,\sigma_A,\sigma_B$, should it be inserted in $\AA$. Figure \ref{fig2} shows possible $v_A,\epsilon$ combinations for a given
molecule mass.
 \\ \\
 The second strategy to diminish the unwanted contribution requires that the molecule rotation is in a chaotic regime. This
means one is allowed to ignore the actual angle trajectories and to average over their distribution. We assume a homogeneous one and
just substitute $|sin(\beta)|,|cos(\beta)|\rightarrow 2/\pi$ etc. . The ratio of unwanted to wanted contribution is then in the order
of $(Q_>-Q_<)\hbar v_A/(4W)$. If we insert $1\AA^{-1}$ as estimate for the momentum cut off (about 10 times the inverse size of a buckyball), the ratio becomes
small if
\beq
97~W\gg v_A
\eeq
is fulfilled - provided one uses the same units as in equation (\ref{inequationwithnumbers}).
Equation (\ref{mainresult1}) has of course to be fulfilled simultaneously, but figure \ref{fig2} exemplifies that this is possible in a realistic
parameter regime.
\begin{figure}[t]
\begin{center}
\epsfig{file=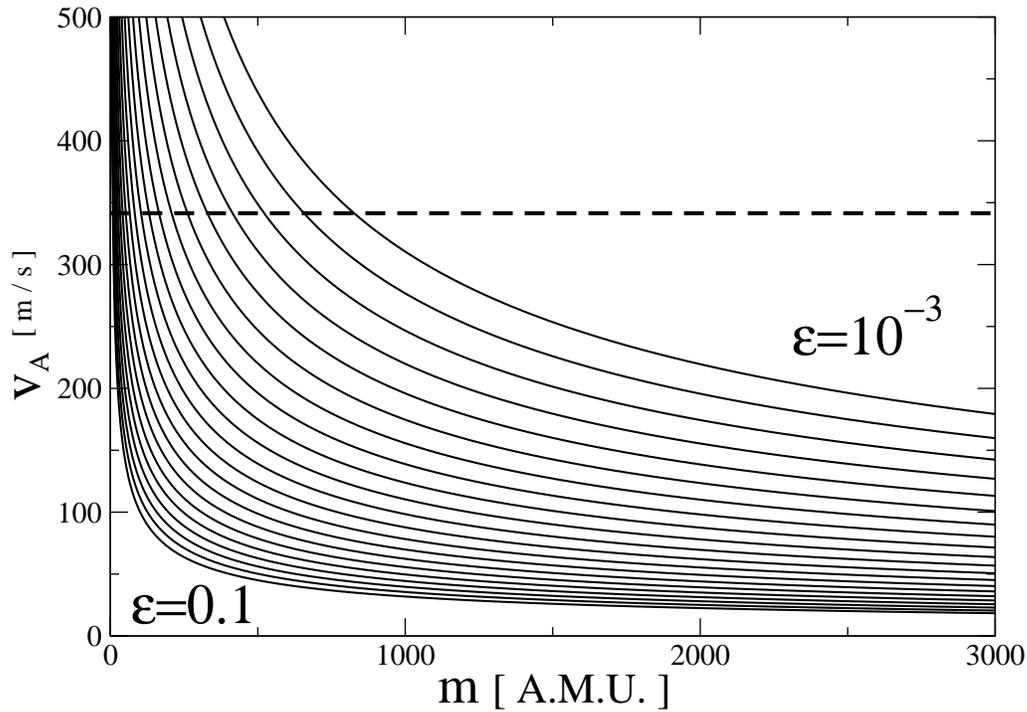, angle=0,clip=true,width=0.85\linewidth}
\end{center}
  \caption{Solutions for equation (\ref{inequationwithnumbers}) are shown for various relative momentum loss $\epsilon$. The dashed line gives the upper limit $Mv_{limit}$ for
the initial velocity $v_A$, the lower limit is far below the lowest shown line.  We have chosen $1meV$ for the spin wave gap and a tube with 10000 rings, take in each ring a nearest neighbour distance of $\rho=3\AA$  
a tube diameter of $100\AA$ - thus each ring contains about 105 coupled spins. 
The relative momentum loss is taken as $\epsilon=10^x$ where the upper line corresponds to $\epsilon=10^{-3}$ and the lowest one to $\epsilon=0.1$,
 $x$ changes from curve to curve by 0.1. The molecule mass is given in atomic mass units (A.M.U.). }
  \label{fig2}
\end{figure}

\section{just an almost good measurement ?\label{chptnoencoding}}
This section should give more quantitative arguments to the discussion in the context of equation (\ref{almostgood}). We elaborate how an
{\it almost} good measurement can become a {\it really good} measurement.\\

We have to understand more about the state in which a spin chain is left after $\stateuu,\statedd$ passed by. It is much more difficult
to calculate these states than just obtaining the transmission amplitudes in the last chapter, because one has to follow the full quantum
dynamics. We show now that it is for the current purpose not necessary. \\

We restrict ourself to the spin ring at $r_n=0$ and use the notation $\ket{\lambda}\equiv\gamma_{\lambda,0}^{\dagger}\ket{GS}$ for the excitable magnon states. 
In the same
spirit as we were using the Born Approximation, we can project out of the Hamiltonian $H_0^{x}+V_{dip}^{x}$ the states with more than one
magnon. The effective Hamiltonian is then in bra-ket notation
\beq
H_{eff}&=&\sum_{k}E(k)\ket{k,GS}\bra{k,GS}\nonumber\\
&&+\sum_{k,k',\lambda}\delta_{k,k'}(E(k)+W)\ket{k,\lambda}\bra{k,\lambda}\pm V_{tot}^{\lambda,0}(k-k')\ket{k,\lambda}\bra{k',GS}~+~h.c
\eeq
where the '+' corresponds to the $\stateuu$- and the '-' to the $\statedd$-channel. The evolutions equal under the transformation
$\ket{\lambda}\rightarrow-\ket{\lambda},\lambda=1,..,4$. An initialised state 
$(\mu_{\uparrow\uparrow}\stateuu+\mu_{\downarrow\downarrow}\statedd)\ket{GS}$ evolves then 
into $(1-|Z|^2)^{-1/2}\left( \mu_{\uparrow\uparrow}\stateuu-\mu_{\downarrow\downarrow}\statedd \right)\ket{ex} 
+ Z\left(   \mu_{\uparrow\uparrow}\stateuu+\mu_{\downarrow\downarrow}\statedd   \right)\ket{GS}$. Where $\ket{ex}$ is some excited state
of the spin ring and $|Z|\leq1$. If this state decays independently of the molecular state, it provides just an arbitrary phase $\Omega$ and
in the final state
\beq
\left[
\mu_{\uparrow\uparrow}
\left(Z+\sqrt{1-|Z|^2}exp(i\Omega))\right)
\stateuu  \right. \nonumber \\
+\left.
\mu_{\downarrow\downarrow}
\left(Z-\sqrt{1-|Z|^2}exp(i\Omega))\right)
\statedd
\right]\ket{GS}
\eeq 
the spin ring factors out. Thus in the limit $|Z|\rightarrow1$, the relative strength of the coefficients is not even changed by the scattering.
The crucial aspect in the argument is that the decay happens independently of the molecular states. Having a longer relaxation time than the
typical time which the molecule stays inside the influence of a single chain should be a sufficient condition for its validity. 
 
\section{summary}
We have proposed a new type of experiment to measure the squared total magnetisation of a molecule in the direction of its spin
anisotropy axis. We have shown in which parameter regimes one can realize it. 
The experimental side meets several challenges. The first key problem is to find an appropriate molecule. It should have the property, that it carries the basis
to create the Bell states by two spins and that the individual spin multiplets have big energy splittings. It is further necessary
that one is able to build a 'magnetic oven' - a source of these molecules, where they get excited into superpositions of these states.
The second big question is the actual realization of the tube. It is a nontrivial goal to fabricate such a mesoscopic object especially
with small diameter but still many individual chains. But we believe it is a way to go, because improved technology can help to make
use of the peculiarities in Quantum Mechanics. 
\acknowledgments
The author thanks J.M.J. van Leeuwen and C.W.J. Beenakker for discussions. This work was supported by the Dutch Science Foundation NWO/FOM. 
\appendix
\section{spin wave approximation}
The spin chain is described by the antiferromagnetic heisenberg model
\beq
H_{AF}\!=J\!\!\!\!\!\!\!\!\!\sum_{\varphi=1/N,2/N,..}^1 \!\!\!\!\!\!\!\!\!\!  \vec{S}_A(\varphi)\vec{S}_B(\varphi)\!+\!
\vec{S}_A(\!\varphi+\!1/N)\vec{S}_B(\varphi)+\!S\Delta(S^z_A(\varphi)\!-\!S^z_B(\varphi))
\eeq
, where $\varphi$ labels the unit cells, $A,B$ stands for the sub-lattices and $\Delta$ mimics e.g. single ion anisotropy and 
stabilises the Neel state. Through the transformation 
\beq
&&S^z_A(\varphi)=a_{\varphi}^{\dagger}a_{\varphi}-S~,~~ S^z_B(\varphi)=S-b_{\varphi}^{\dagger}b_{\varphi}\\
&&S^+_A(\varphi)=a^{\dagger}_{\varphi}\sqrt{2S-a^{\dagger}_{\varphi}a_{\varphi}}~,~~
S^+_B(\varphi)=\sqrt{2S-b^{\dagger}_{\varphi}b_{\varphi}}b_{\varphi}\\
&&S^-_A(\varphi)=\sqrt{2S-a^{\dagger}_{\varphi}a_{\varphi}}a_{\varphi}~,~~
S^-_B(\varphi)=b^{\dagger}_{\varphi}\sqrt{2S-b^{\dagger}_{\varphi}b_{\varphi}}\\
&&a_{\varphi}=N^{-\frac{1}{2}} \!\!\!   \sum_{k=0,1/N,..}^{(N-1)/N} \!\!\!\!\!\!   e^{i2\pi k(\varphi+\frac{1}{2N})}\left(cosh(\Theta_k)\alpha_k+sinh(\Theta)\beta_k^{\dagger}  \right)\\
&&b_{\varphi}=N^{-\frac{1}{2}} \!\!\! \sum_{k=0,1/N,..}^{(N-1)/N}e^{-i2\pi k\varphi}\left(cosh(\Theta_k)\beta_k+sinh(\Theta)\alpha_k^{\dagger}  \right)\\
&&tanh(2\Theta_k)=\frac{cos(k\pi)}{1+\Delta/2S}
\eeq
,referenced to Holstein and Primakoff, is a mean field hamiltonian 
\beq
H_{AF}&=&-2JNS(S+1)-JN(2S+1)\Delta\nonumber\\
&&~+\sum_k \omega_k\left(\alpha_k^{\dagger}\alpha_k+\beta_k^{\dagger}\beta_k+1\right)+O(S^0) \\
\omega_k&=&JS\sqrt{(2+\Delta/S)^2-4cos^2(k\pi)}
\label{meanfieldh}
\eeq
obtained, which contains the decoupled magnons. It corresponds formally to a 
large S expansion around the Neel Groundstate.
\section{magnetic dipole interaction}
 The magnetic dipole interaction between the 
spins of the molecule and the chain (in the plane with $z=0$) should now be expressed in the same approximation.
\beq
V_{dip}(\vec{R})\!&=&\!\mu_0\mu_e^2\!\sum_{\varphi;\chi;i}
\frac  {\vec{\sigma}_i\vec{S}_{\chi}(\varphi)}    {|\vec{R}(\varphi)|^{3}}
-3\frac  {   \left[\vec{R}(\varphi)\vec{\sigma}_i\right]\left[\vec{R}(\varphi)\vec{S}_{\chi}(\varphi)\right]  }    {|\vec{R}(\varphi)|^5} \\
\vec{R}(\varphi)&=&\rb{d~cos(2\pi\varphi/N),d~sin(2\pi\varphi/N),R}
\eeq
contains the magnetic constant of the vacuum $\mu_0$, the electron spin moment $\mu_e$, the molecular z component, its spin degrees
$\sigma_i$.~$\chi$ labels the $A,B$ sub-lattices. d is the tube diameter. The molecular
spins are quenched along the axis in direction
$\vec{l}=(cos(\beta)cos(\alpha),cos(\beta)sin(\alpha),sin(\beta))$. Thus one
obtains for the effective interaction $\bra{x}V_{dip}(\vec{R})\ket{x'}=\delta_{x,x'}V^x_{dip}(R)$
\beq
V^{\stateud}_{dip}(R)=V^{\statedu}_{dip}(R)&=&0 \\
V^{\stateuu}_{dip}(R)=-V^{\statedd}_{dip}(R)&\equiv&V_{asym}+V_{sym}
\eeq
\beq
 V_{asym}&=&\frac{-3dR\mu_0\mu_e^2}{|\vec{R}(\varphi)|^5}
\sum_{\chi,\varphi}
\left(
sin(\beta) cos(2\pi\frac{\varphi}{N}) S^{x}_{\chi}(\varphi)
+sin(\beta)  sin(2\pi\frac{\varphi}{N}) S^y_{\chi}(\varphi) \right.\nonumber \\
 &&\left. +cos(\beta)cos(\alpha+2\pi\frac{\varphi}{N})  S^z_{\chi}(\varphi)
  \right) \nonumber
\eeq
\beq
V_{sym}=\nonumber \\
\frac{\mu_0\mu_e^2}{|\vec{R}(\varphi)|^3}
\sum_{\chi,\varphi}
\left(
cos(\beta)cos(\alpha)S_{\chi}^x(\varphi)
+cos(\beta)sin(\alpha)S_{\chi}^y(\varphi)
+sin(\beta)S_{\chi}^z(\varphi) 
\right)
 \nonumber \\
+\frac{-3d^2\mu_0\mu_e^2}{|\vec{R}(\varphi)|^5}
\sum_{\chi,\varphi}
\left(
cos(\beta)cos\left(\alpha+2\pi\frac{\varphi}{N}\right)cos(\varphi)\right.\nonumber\\
\left. \times S_{\chi}^x(\varphi)
+cos(\beta)cos(\alpha+2\pi\frac{\varphi}{N})sin(\varphi)S_{\chi}^y(\varphi)
+sin(\beta)S_{\chi}^z(\varphi)
\right).\nonumber \\
&&
\eeq
We have split the parts that are symmetric and antisymmetric under $R\rightarrow-R$. If one expresses
the spins by the magnon operators one finds, that the interaction couples only to four degenerate modes
\beq
\vec{\gamma}&\equiv&\{\alpha_{1/N},\beta_{1/N},\alpha_{1-1/N},\beta_{1-1/N}  \}
\label{fourmodes}
\eeq
with the energy of the anisotropy gap $W=2J\Delta\sqrt{1+\Delta/S}+{\cal O}(1/N)$  
\beq
 V_{asym}=\sqrt{\frac{SN}{2}}sin(\beta)
\frac{3Rd\mu_0\mu_e^2}{\left(R^2\!+\!d^2\right)^{5/2}}
\left[
cosh(\theta_0)(\gamma_1^{\dagger}-\gamma_4^{\dagger})
+sinh(\theta_0)(\gamma_2^{\dagger}+\gamma_3^{\dagger})
 \right]\nonumber \\
+h.c.~+{\cal O}(1/N)
\eeq
\beq
 V_{sym}=\mu_0\mu_e^2\sqrt{\frac{SN}{2}}cos(\beta)\left(cosh(\theta_0)+sinh(\theta_0)\right) \nonumber \\
\times
\left[
\frac{e^{-i\alpha}\gamma_1^{\dagger}+e^{i\alpha}\gamma_2^{\dagger}}{(R^2+d^2)^{3/2}}
-\frac{3d^2}{2(R^2+d^2)^{5/2}}
\left(
e^{i\alpha}\gamma_1^{\dagger}
+e^{-i\alpha}\gamma_2^{\dagger}
-e^{-i\alpha}\gamma_3^{\dagger}
+e^{i\alpha}\gamma_4^{\dagger}
\right)
\right]\nonumber \\
+h.c.~+{\cal O}(1/N).
\eeq
It is desirable to express now all in the second quantised form
\beq
 \quad V_{dip}^{\stateuu}=
\sum_{\lambda,n}\int dkdk`
a_k^{\dagger}a_{k`}\gamma_{\lambda,n}^{\dagger}
[
V_{s}^{\lambda,n}(k-k`)
+V_{a}^{\lambda,n}(k-k`)
]
+h.c.
\eeq
, in which $(\lambda,n)$ labels the $\lambda$-th of the four modes in the chain with z coordinate $r_n$.  
The fourier transform lead to the modified Besselfunctions of second kind $B[..]$
\beq
 V_{a}^{1,n}(q)=ie^{-ir_nq}\mu_0\mu_e^2\sqrt{\frac{SN}{2}}sin(\beta)cosh(\theta_0)q^2B[1,|qd|] \nonumber
\eeq
\beq
V_{a}^{2,n}(q)=V_{a}^{3,n}(q)=tanh(\theta_0)V_{a}^{1,n}(q)
\mbox{ and }
V_{a}^{4,n}(q)=-V_{a}^{1,n}(q)&&
\eeq 
\beq
V_{s}^{1,n}(q)=V_{s}^{2,n}(q)^*&=&e^{-ir_nq}\mu_0\mu_e^2\sqrt{\frac{SN}{2}}cos(\beta)
\left(cosh(\theta_0)+sinh(\theta_0)\right) \nonumber \\
&&\times
\left[
2e^{-i\alpha}\left| \frac{q}{d} \right|B[1,|qd|]
-e^{i\alpha}q^2B[2,|qd|] 
\right] \nonumber
\label{Vasym}
\eeq
\beq
V_{s}^{3,n}(q)=-V_{s}^{4,n}(q)^*&=&e^{-ir_nq}\mu_0\mu_e^2\sqrt{\frac{SN}{2}}cos(\beta)
\left(cosh(\theta_0)+sinh(\theta_0)\right)\nonumber \\
&&\times e^{-i\alpha}q^2B[2,|qd|]
\label{Vsym}
\eeq


\section*{References}
\begin{thebibliography}{99}
\bibitem{qinf1}
Preskill J, Lecture notes on Quantum Computation, Course 219, 
{\it http://www.theory.caltech.edu/people/preskill/ph229/}
\bibitem{qinf2}
Nielsen M A,  Chuang I L,
   {\it Quantum Computation and Quantum Information},2000 (Cambridge Univ. Press, New
York)
\bibitem{parityidea}
Beenakker C W J, DiVincenzo D P, Emary C, and Kindermann M
,Charge detection enables free-electron quantum computation, 2004
  ,{\it Phys.
  Rev. Lett.} {\bf 93} 020501
\bibitem{parityidea2}
Carlos Egues J, Fingerprinting Spin Qubits, 2005,
 {\it Science} {\bf 309}, 565

\bibitem{parity1}
Wenjin M, Averin D V, Ruskov R and  Korotkov A N,
Quadratic Quantum Measurements, 2004
 {\it Phys. Rev. Lett.} {\bf 93}, 056803
\bibitem{parity2}
Engel H A and Loss D,
Fermionic Bell-State Analyzer for Spin Qubits, 2005,
 {\it Science} {\bf 309}, 586
\bibitem{parity3} 
Trauzettel B, Jordan A N, Beenakker C W J , Buttiker M, 
 Parity meter for charge qubits: an efficient quantum entangler , 2006,
{\it Phys. Rev. B} {\bf 73}, 235331
\bibitem{parity4}
Schomerus H and Robinson J P, Entanglement between static and flying qubits in an
Aharonov-Bohm double electrometer, 2006, {\it arXiv:cond-mat/0604498}

\bibitem{field}
Field M, Smith C G, Pepper M, Ritchie D A, Frost J E F , 
Jones G A C and Hasko D G,
Measurements of Coulomb blockade with a noninvasive voltage probe
, 1993,
{\it Phys. Rev. Lett.} {\bf 70}, 1311

\bibitem{encoding}
Titov M, Trauzettl B, Michaelis B and Beenakker C W J,
Transfer of entanglement from electrons to photons by optical selection rules, 
{\it New Journal of Physics} {\bf 7}, 186

\bibitem{NegeleOrland}
Negele J W, Orland H, {\it Quantum Many-Particle Systems}, 1988,
Addison-Wesley Publishing Company,New York

\bibitem{holsteinprimakoff}
Holstein T and Primakoff H, Field Dependence of the Intrinsic Domain Magnetization of a Ferromagnet, 1940,
{\it Phys. Rev.} {\bf 58}, 1098

\end{thebibliography}
\end{document}